# High Coherence Mid-Infrared Dual Comb Spectroscopy Spanning 2.6 to 5.2 microns

Gabriel Ycas,[1*] Fabrizio R. Giorgetta,[1] Esther Baumann,[1] Ian Coddington,[1] Daniel Herman,[1] Scott A. Diddams,[1] Nathan R. Newbury[1]

[1]*NIST 325 Broadway, Boulder, CO 80305*
*\*Corresponding author: ycasg@colorado.edu*

Mid-infrared dual-comb spectroscopy has the potential to supplant conventional high-resolution Fourier transform spectroscopy in applications that require high resolution, accuracy, signal-to-noise ratio, and speed. Until now, dual-comb spectroscopy in the mid-infrared has been limited to narrow optical bandwidths or to low signal-to-noise ratios. Using a combination of digital signal processing and broadband frequency conversion in waveguides, we demonstrate a mid-infrared dual-comb spectrometer that can measure comb-tooth resolved spectra across an octave of bandwidth in the mid-infrared from 2.6—5.2 μm with sub-MHz frequency precision and accuracy and with a spectral signal-to-noise ratio as high as 6500. As a demonstration, we measure the highly structured, broadband cross-section of propane ($C_3H_8$) in the 2860-3020 cm$^{-1}$ region, the complex phase/amplitude spectrum of carbonyl sulfide (COS) in the 2000 to 2100 cm$^{-1}$ region, and the complex spectra of methane, acetylene, and ethane in the 2860-3400 cm$^{-1}$ region.

Mid-infrared spectroscopy is a powerful technique for the multispecies detection of trace gases with applications ranging from the detection of hazardous materials, to environmental monitoring and industrial monitoring. Compared to the near-infrared, where laser sources are more plentiful, the techniques for measuring mid-infrared spectra are more limited. Mid-infrared spectra are most commonly acquired by Fourier transform spectroscopy (FTS), which provides accurate and high resolution spectra but requires a scanning delay arm and blackbody source leading to large instruments and long acquisition times. Dual-comb spectroscopy (DCS) is a high-performance alternative to conventional FTS providing high resolution, absolute frequency accuracy, fast acquisition times, long interaction lengths, broad bandwidth coverage, and high signal-to-noise ratio [1,2]. The advantages of speed and long path length are of particular relevance to non-laboratory applications, for example in open-path atmospheric monitoring or industrial process monitoring [3–6]. However, up until now, DCS has only been demonstrated with its full panoply of advantages in the near-infrared, from ~ 1 to 2 μm [7–10]. The near-infrared has much more limited applications compared to the mid-infrared since molecular cross-sections are typically 1000 times weaker, if they exist at all. DCS in the mid-infrared has indeed been actively pursued [11–24], yet it is not competitive with high-resolution conventional FTS, limited by the coherence and/or the bandwidth of mid-infrared comb sources.

Quantitative broadband mid-infrared DCS requires addressing strong overlapping requirements on the underlying mid-infrared frequency combs: they must produce broad and relatively flat optical spectra while maintaining mutual coherence over the measurement time. Without coherence, adjacent comb teeth blend together, sacrificing orders of magnitude in spectral resolution and obscuring both the frequency and amplitude accuracy possible with this technique. The requirements for bandwidth and coherence are strongly coupled -- as the optical bandwidth increases (or required SNR increases), the measurement time must correspondingly increase, putting ever more stringent requirements on the mutual coherence between the combs. Driven by the challenges involved in generating highly coherent mid-infrared light, high signal-to-noise ratio (SNR), accurate and quantitative mid-infrared DCS have so far been limited to narrow instantaneous bandwidths [13,17,20].

In this Article, we demonstrate mid-infrared dual-comb spectroscopy with high coherence, recording a high-fidelity spectrum of the complex molecule propane. We capture both the broad 150 cm$^{-1}$ pedestal and complicated fine structure of this gas, which has features as narrow as 0.01 cm$^{-1}$ (300 MHz). We demonstrate, for the first time that, high coherence DCS spectral of large molecules can have equivalent fidelity to high-resolution FTS spectra while being acquired the spectra in 1/75$^{th}$ the time and without requiring correction for instrument lineshape or frequency calibration. We also measure the spectrum of a gas mixture of methane, acetylene, carbonyl sulfide, ethane, and water, measuring the complex (phase and amplitude) spectra in both the 3-micron region and 5-micron region, spanning the entire L and M atmospheric transmission windows.

# EXPERIMENTAL SETUP

## Optical system

Our mid-infrared laser frequency combs use difference frequency generation (DFG) to convert near-infrared light into 2600-5200 nm mid-infrared light, as shown in Figure 1. Each system is based upon a self-referenced polarization-maintaining fiber frequency comb [28] with a repetition rate of 200 MHz, locked to a common 1560 nm cw laser. The near infrared comb output is preamplified to about 30 mW and then split equally into three branches. The first branch generates the f-2f signal used for locking the comb's offset frequency. The second branch generates pump light for the DFG process at 1070 nm. The preamplifier output is amplified in a core-pumped erbium-doped fiber amplifier (EDFA) and launched into a highly nonlinear optical fiber (HNLF) with anomalous dispersion. This provides about 1 mW of power between 1040-1090 nm, which is filtered, stretched with a chirped fiber Bragg grating, amplified to 1.6 W in a ytterbium-doped fiber amplifier, and compressed with a Treacy compressor to ~200 fs and 900 mW of average power. The third branch generates signal light spanning 1350 to 1750 nm by amplification in a core-pumped EDFA to 430 mW and subsequent broadening in normal dispersion HNLF.

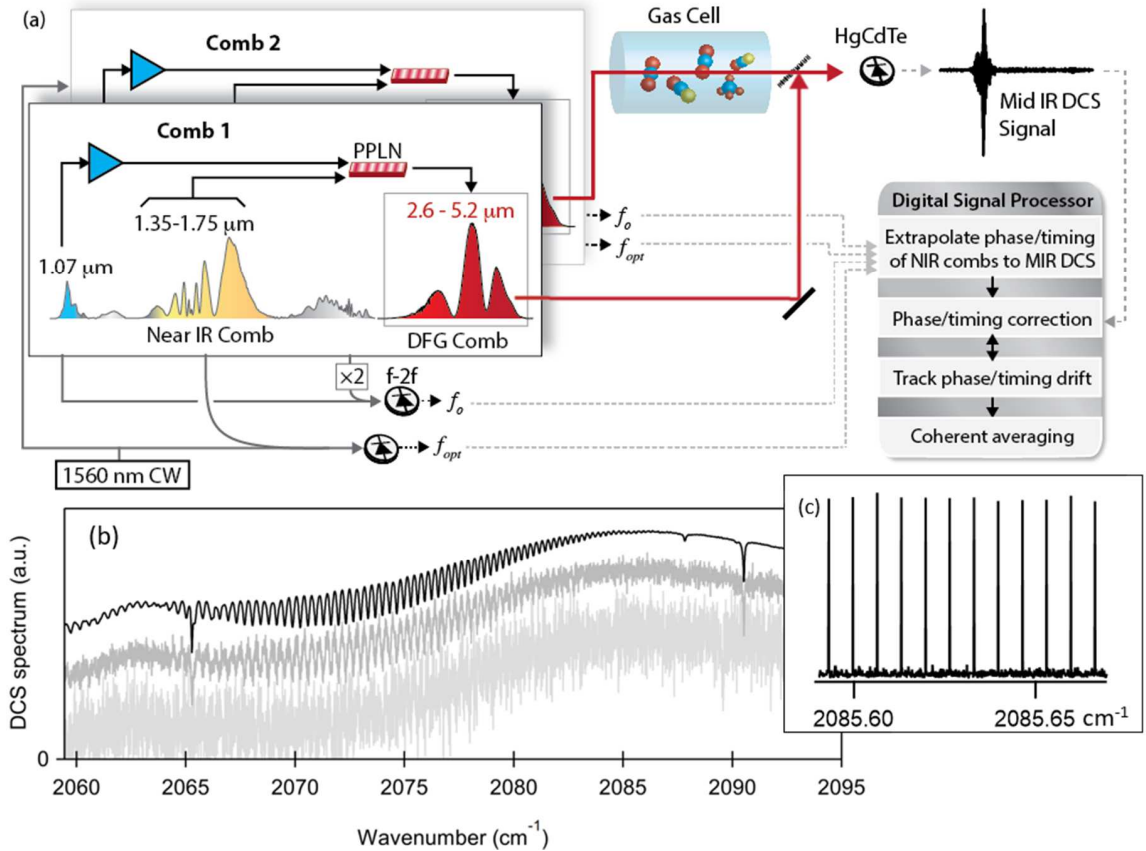

Figure 1. **Octave-spanning mid infrared dual comb spectrometer**. (a) Near infrared laser frequency combs generate mid-infrared light by difference frequency generation (DFG) in bulk or chirped waveguide periodically-poled lithium niobate (PPLN). Light from one comb is passed through a gas cell and combined with the second comb to generate a series of interferograms that are detected on a mercury-cadmium-telluride (HgCdTe) photodetector. A digital signal processor corrects the phase and timing of the detected interferograms based on the instantaneous near-infrared combs' offset frequencies, $f_o$, and optical comb tooth frequencies near 1560 nm, $f_{opt}$, as well as slower environmentally-driven phase/timing drifts from the other optical subsystems. The Fourier transform of these corrected interferograms yields the mid-IR spectrum. (b) Example mid-IR spectrum of carbonyl sulfide (COS) near 5 μm (2000 cm$^{-1}$) at coherent averaging times of 10 ms (light gray), 100 ms (dark gray), and 100 s (black), illustrating the expected improvement in SNR with coherent averaging time. (c) Example comb-tooth resolved spectrum in the same spectral region for 600-ms of streamed phase/timing corrected DCS data.

Light from the pump and signal branches is combined on a dichroic mirror and focused into a PPLN crystal using an off-axis paraboloid mirror. We have used both a bulk PPLN crystal 1 mm in length [14,17,19,26,27] and a chirped PPLN waveguide. The chirped waveguide PPLN generates a structured but broad spectrum with an instantaneous bandwidth exceeding 1600 cm$^{-1}$.; its design is discussed in Section 3.C. The bulk crystal provides smoother spectra with optical bandwidths from 760 to 120 cm$^{-1}$ (-10 dB) as the center wavelength is tuned from 3 to 5 μm by selecting different poling periods, as shown in Fig. 2. The output powers range from 40-100 mW for wavelengths from 2.5 – 3.8 μm and 1-10 mW for wavelengths from 4 to 5 μm. These powers are sufficient to support future open-path measurements; for the laboratory data here, the mid-infrared light is attenuated to a few hundred microwatts to avoid saturation of the HgCdTe detector.

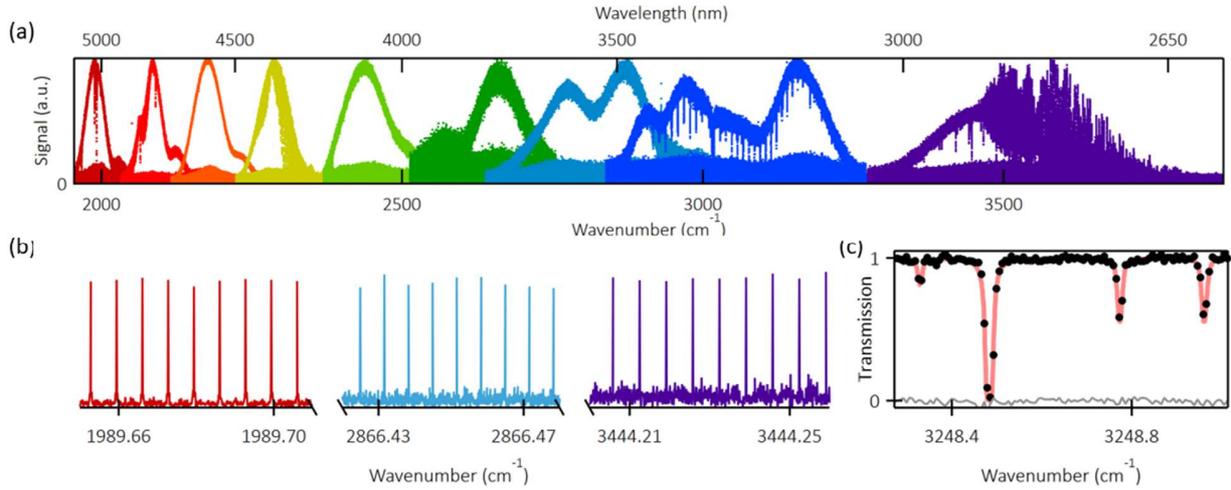

Figure 2: **Nine dual-comb spectra** (a) at different PPLN poling periods, for ~0.6 second of streamed, phase and timing corrected DCS data. The spectra span ~1800 cm$^{-1}$ bandwidth or 270,000 resolved comb modes. They have been normalized to equal peak height. The structure above 3500 cm$^{-1}$ is due to water absorption in ambient air. (b) Expanded view showing resolved comb modes spaced by 0.0067 cm$^{-1}$ (200 MHz) across the covered spectral range. These spectra are shown before coherent averaging which only retains the values at the DCS comb teeth. (c) Expanded view of a coherently averaged DCS spectrum for a 2.6 Torr cell of HCN. The DCS is sampled at 200-MHz comb tooth spacing, as indicated by the black dots, and falls directly on the HCN spectra from refs. [32,33] (red line).

The choice of DFG for the mid-infrared combs is not the only option. In addition to DFG, other mid-infrared DCS demonstrations have emphasized novel sources such as quantum cascade lasers [23,28], femtosecond lasers [12], micro resonators [21], and optical parametric oscillators [15,18,22,24]. The DFG combs have a combination of attributes that are attractive. First, the combs' phase noise can be easily measured in the near-infrared and then extrapolated to the mid-infrared region. Second, the DFG combs can provide powers up to 100 mW, more than sufficient for operation in stressing, long-path configurations. Third, the DFG process is relatively simple so that the combs operate continuously for many days. Finally, since the entire mid-infrared DCS optical system easily fits on a 3'x5' table, it can be made portable for non-laboratory use where the improvements of mid-infrared DCS over conventional high resolution FTS would have the most impact.

**Data acquisition and processing for comb tooth-resolved spectra**

Multiple approaches have been taken to achieve mutual comb coherence needed for dual-comb spectroscopy, but often at the cost of bandwidth or experimental complexity [1]. Of these approaches, digital signal processing alone can avoid these tradeoffs while enforcing not only mutual coherence but also enabling the coherent averaging needed to acquire high SNR spectra. Real-time phase correction through digital signal processing has only been so far demonstrated in a relatively narrow optical region around a pair of continuous-wave lasers acting as transfer-oscillators [7,29]. Here we generalize the digital signal processing approach to support broadband DCS by allowing for the correction of a DCS signal at any color using an arbitrary pair of optical references. In our system, this is demonstrated by our use of two near-infrared signals to correct a mid-infrared DCS signal, but the approach is general.

We require the underlying combs obey the standard comb equation, where the phase of the $n^{th}$ mode is

(1) $$\theta_n(t) = n\theta_r(t) + \theta_0(t),$$

where $\theta_r(t) = 2\pi f_r t + \delta\theta_r(t)$ and $\theta_0(t) = 2\pi f_0 t + \delta\theta_0(t)$ in terms of the usual repetition frequency $f_r$, and offset frequency $f_0$, and where we allow for phase noise on the comb's repetition rate and offset frequency, $\delta\theta_r(t)$ and $\delta\theta_0(t)$. In DCS, many pairs of modes are simultaneously mixed on a square-law photodetector to generate a series of DCS radio-frequency modes, each with phase

(2) $$\phi_k(t) = k\phi_r(t) + \phi_0(t),$$

where $k$ is the index of a radio-frequency DCS mode and

$$\phi_r(t) = 2\pi\Delta f_r t + \Delta\delta\theta_r(t)$$
$$\phi_0(t) \approx 2\pi\Delta f_0 t + n_0\phi_r(t) + \Delta\delta\theta_0(t) - 2\pi m f_{r,comb1}t,$$

**(3)**

where

$$\Delta f_r = f_{r,comb2} - f_{r,comb1}, \quad \Delta\delta\theta_r = \delta\theta_{r,comb2} - \delta\theta_{r,comb1},$$
$$\Delta f_0 = f_{0,comb2} - f_{0,comb1}, \quad \Delta\delta\theta_0 = \delta\theta_{0,comb2} - \delta\theta_{0,comb1}.$$

$n_0$ is the optical comb-mode index for comb 2 corresponding to the rf DCS mode when $k=0$. The difference in repetition rates is chosen such that $f_{r,comb1}/\Delta f_r$ is an integer and $m \equiv n_0\Delta f_r/f_{r,comb1}$ is present because a DCS signal is typically measured at radio frequencies between 0 and $f_r/2$ and in many cases $n_0\Delta f_r > f_r/2$. Thus, the measured signal can be the heterodyne of modes $n$ from comb 2 with modes $n+m$ from comb 1.

The DCS spectrum loses coherence if the phase noise on its $k^{th}$ mode, $\delta\phi_k(t) = (n_0 + k)\delta\Delta\theta_r(t) + \delta\Delta\theta_0(t)$, exceeds a radian, leading to a loss in signal-to-noise ratio as $\exp(-\langle\delta\varphi_k^2\rangle)$. Moreover, if the phase noise extends across neighboring modes, the comb tooth resolution is lost and, with it, the frequency resolution and accuracy that distinguishes DCS from FTS. To avoid this decoherence, we correct the DCS interferogram signal in the time domain, assuming knowledge of $\delta\Delta\theta_r$ and $\delta\Delta\theta_0$. First, we rotate the phase of the digitized interferogram by $-\phi_0(t)$. Second, the timing variations are removed by resampling the recorded interferogram by linear interpolation onto a new time axis, determined by

**(4)**
$$t' = t + \Delta\delta\theta_r(t)/2\pi f_r.$$

To perform these corrections, it is necessary to measure $\theta_r(t)$ and $\theta_0(t)$ for each mid-infrared comb with sufficient precision to evaluate $\delta\phi_k(t)$ to much better than one radian. Our mid-infrared light is generated using difference-frequency generation, and consequently the mid-infrared combs have $f_0 = \delta\theta_0 = 0$. However, we must measure $\Delta\delta\theta_r$ with sufficient precision such that $(n_0 + k)\Delta\delta\theta_r$ is much less than one radian at short (tens of microsecond) timescales. To reach this precision via direct measurements of the mid-infrared combs' repetition-rate signal is extremely demanding and has not yet been demonstrated. Instead, we extract this quantity from measurements of the near-infrared combs as follows. The near-infrared $\theta_0(t)$ is measured using an f-2f interferometer. A second measurement of each near-infrared comb's phase is made by beating the $n_{opt}$ mode near 1560 nm against a common cw laser, providing a phase signal $\delta\theta_{opt}(t)$. The difference $\Delta\delta\theta_{opt}(t) \equiv \delta\theta_{opt,comb2}(t) - \delta\theta_{opt,comb1}(t)$ is calculated, removing dependence on the cw laser's phase, and finally we calculate $\delta\Delta\theta_r(t) = n_{opt}^{-1}[\Delta\delta\theta_{opt}(t) - \Delta\delta\theta_0(t)]$.

In an extension of earlier work [29,7], we implement the phase-correction algorithm using standard signal processing techniques in real time on a field programmable gate array (FPGA.) The radio-frequency signals from each of the near infrared combs' two locks – in our case $f_0$ beat notes (via self-referencing) and at 1560 nm (against a cw laser) -- and the interferogram are digitized with a 14 bit analog-to-digital converter that is clocked on comb 1's repetition rate. These real signals are converted to complex ones using IQ demodulation, and the phase errors are calculated using the arctangent operation implemented using a CORDIC. From these phase error signals, point-by-point phase and timing corrections to the mid-infrared DCS signals are computed from the above equations using fixed-point arithmetic and applied to the DCS signal by multiplication with a complex phasor and linear interpolation, respectively. Here, the near infrared frequency combs are sufficiently tightly phase-locked that the timing resampling is unnecessary, provided that the phasor is computed not by $\phi_0(t)$, but instead by $\delta\phi_k(t)$, where the mode $k$ is centered in the mid-infrared DCS spectrum.

This high-bandwidth, real-time phase correction compensates for phase noise originating with the near infrared combs. To compensate for additional slow drifts in timing and phase from optical path length variations in the amplification, DFG setup, and optical cell, a cross-correlator is implemented on the FPGA which compares the stream of interferograms to a reference interferogram. The position and phase of the cross-correlation peaks are used as error signals for proportional-integral feedback to the real-time phase and timing correction processor [7], which "locks" the timing and phase offsets to zero on a timescale of about 10 interferograms, here compensating for noise slower than 10 Hz.

In addition to the real-time phase correction, the FPGA signal processor also implements coherent averaging; coherent averaging is critical to reducing the data size for any reasonable averaging time [30]. A single interferogram, up to ~8 million points long, is buffered in dynamic random access memory (DRAM.) After phase/timing corrections, subsequent interferograms are added point-by-point to this interferogram and written back to the buffer. This process can be repeated to average up to 65,000 interferograms on the FPGA, at which point they are transmitted to the PC over the RIFFA PCIe interface [31].

Figure 2 shows a series of coherent, comb tooth-resolved spectra for tunings of the PPLN from 2.6—5.2 μm. For each spectrum, 69 interferograms with 1,922,334 samples each were recorded at $\Delta f_r \approx 104$ Hz in 0.66 seconds and then Fourier transformed to produce optical spectra. We observe resolved DCS comb modes with no signs of line broadening at all spectral bands, demonstrating that for all mid-infrared spectra the relative comb linewidth is less than 2 Hz. For longer acquisitions we implement coherent averaging as described above, which effectively retains only the (complex) values at the DCS rf comb teeth, to generate high SNR spectra as shown in Figure 1b and in the remainder of this article. With coherent averaging, the SNR improves as square root of time, essentially indefinitely (see e.g. Fig. 1 (b) and Supplemental Fig. 1). The coherently averaged spectra are sampled at an optical 200 MHz point-spacing where each sample has a frequency accuracy of 100 kHz, set by the fractional uncertainty of the counted repetition rates, and a linewidth less than 1 MHz, determined by the absolute linewidth of the frequency combs. Figure 2 (c) shows a high-resolution spectrum of several well-characterized ~ 300 MHz-wide, HCN absorption lines [32,33].

The digital signal processing system presented here can be applied to almost any dual-comb spectroscopy system, and provides a technically simple means of achieving mutual optical coherence between a pair of combs. Significantly, the signal processing system does not rely on any relationship between the two optical locking frequencies and the frequency of the dual-comb signal – for example, the same signal processor should function equally well with two telecom-band CW lasers and a visible light interferogram.

## RESULTS AND DISCUSSION

### Propane Absorption Cross-section

In contrast to previous DCS demonstrations which have focused on narrow absorption features, we use our dual-comb spectrometer to measure the 140 cm$^{-1}$ wide absorption cross-section of low pressure propane gas. Propane was chosen due to its broad absorption feature at 3.4 μm, which is qualitatively similar to the spectra of other large organic molecules, and also for its relevance to air quality and oil and gas monitoring efforts [34]. Accurately measuring the absorption cross section requires not only high resolution to capture the fine structure, but also a flat spectral response over a wide bandwidth to capture the broader absorption pedestal.

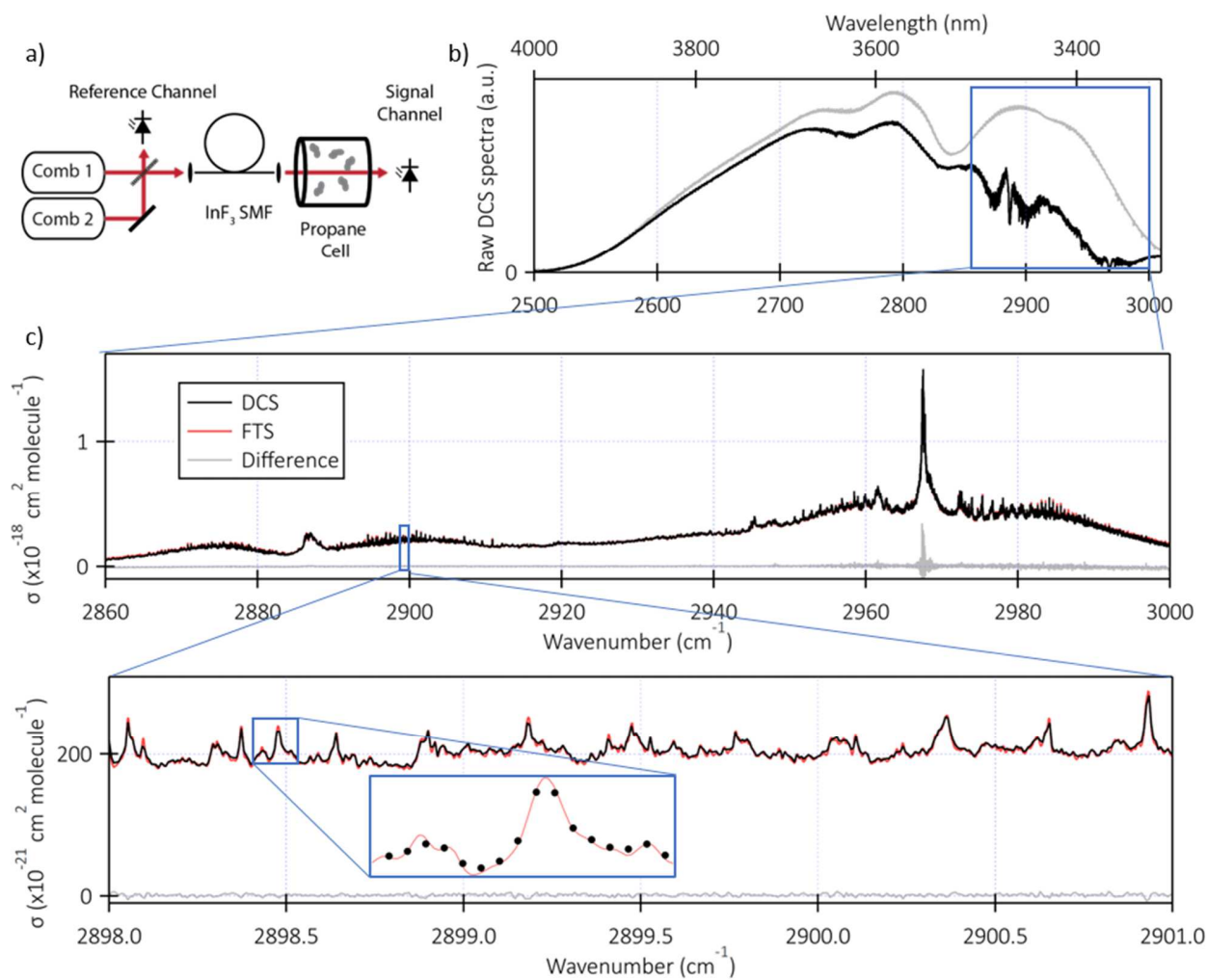

Figure 3. **Propane spectrum and cross section** a) For the propane measurement, the light of both combs is combined on a 50:50 splitter with one output used as reference channel while the second signal channel is sent through a single mode InF$_3$ fiber before probing the propane cell. b) Raw DCS spectrum for the signal channel with the propane cell (black) and without (gray). c) Propane cross-section, σ, extracted from the normalized DCS data. The measured cross-section was shifted by a constant offset to be zero around 2700 cm$^{-1}$ where no absorbers are present. The DCS measured cross-section, spanning 140 cm$^{-1}$ and 21,000 comb modes, compares well to a high-resolution FTS measurement by Beale et al. [35] (red line), with their difference (gray line) flat to within $5\times10^{-21}$ cm$^2$ molecule$^{-1}$ (standard deviation) other than at 2967 cm$^{-1}$ where there is full extinction in the DCS spectrum. d) Detail of the propane cross-section, showing its highly structured nature and excellent agreement between the DCS and high-resolution FTS. As further magnification shows, the width of some of the narrower spikes are around 0.01 cm$^{-1}$, compared to the comb tooth spacing of 0.0067 cm$^{-1}$.

A gas cell was constructed with wedged $CaF_2$ windows to avoid etalon effects and filled with propane gas at low pressure. Since the mid-infrared comb spectra, here generated in a bulk PPLN crystal, change slowly with time, a signal spectrum, $I_{sig}$, and reference spectrum, $I_{ref}$, were measured simultaneously, as shown in Fig. 3 (a). A static discrepancy between them was calibrated out by recording the pair $I_{sig}^{cal}$ and $I_{ref}^{cal}$ without the gas cell. Figure 3 (b) shows the raw spectra $I_{sig}$ and $I_{sig}^{cal}$ spanning 700 cm$^{-1}$ and 100,000 teeth. The peak spectral SNR is 413 for the 4.8 minute-long acquisition, corresponding to $24/\sqrt{s}$. We calculate the normalized transmission as $T_{propane} = \left(I_{sig} I_{ref}^{cal}\right) / \left(I_{ref} I_{sig}^{cal}\right)$ and then the molecular absorption cross section as

$$\sigma = -\ln\left(T_{propane}\right) \cdot \frac{k_B T}{pL}, \quad (6)$$

where the cell length was $L = 13$ cm, the propane pressure was $p = 4.5$ Torr, and the room temperature was $T = 293$ K. As shown in Figure (c-d), the measured cross-section agrees well with a recent high-resolution FTS measurement [35], capturing both a broad absorption envelope and many narrow features. The statistical uncertainty of the cross-section scales inversely with the signal strength of the raw spectra; it is $\sim 5 \times 10^{-21}$ cm$^2$ molecule$^{-1}$ around 2960 cm$^{-1}$ and $\sim 10^{-21}$ cm$^2$ molecule$^{-1}$ around 2900 cm$^{-1}$. Note that at the highest cross-section value at 2967.5 cm$^{-1}$ the DCS measurement is fully saturated, resulting in an increased uncertainty. The SNR for the DCS and FTS measurements is similar but the FTS measurement time is 12 hours, while the DCS spectra were acquired in 9.6 minutes or $\sim$ 75x faster. Moreover, no instrument lineshape correction, baseline correction, or separate frequency axis calibration was needed for the DCS data. These data represent the first quantitative comparison between a broad, structured spectrum measured via conventional high-resolution FTS and via DCS.

### Complex (phase and absorption) spectra of gas mixtures

In a second measurement, the setup was configured as shown in Fig. 1 (a) and the gas cell was filled with a mixture of 5000 ppm methane ($CH_4$), 1000 ppm carbonyl sulfide (COS), 5000 ppm acetylene ($C_2H_2$), and 500 ppm ethane ($C_2H_6$) buffered with nitrogen at atmospheric pressure. Spectra were acquired at two different spectral bands by selecting the poling period of the bulk PPLN crystal.

Figure 4 shows the measured spectrum over 2.9-3.6 μm, corresponding to a bandwidth of 650 cm$^{-1}$ (20 THz) and 100,000 resolved comb modes.

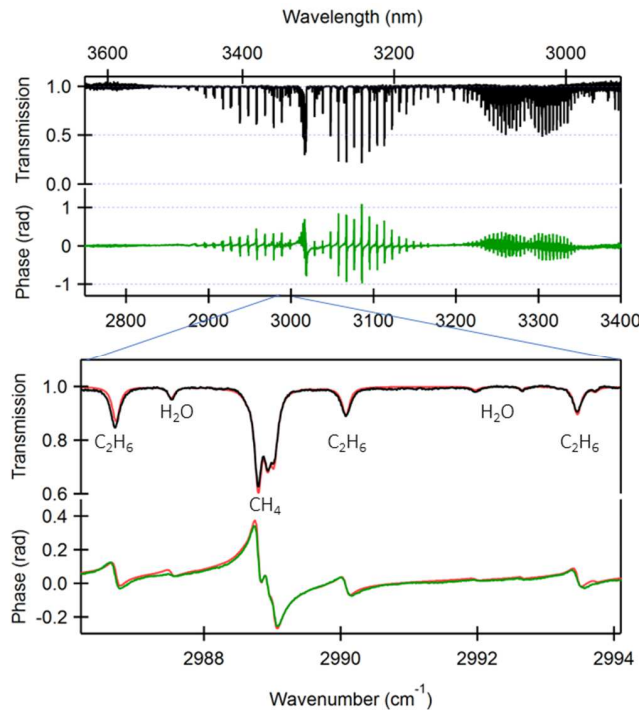

Figure 4. **Dispersive dual-comb spectrum of gas mixture.** Measured spectrum of $CH_4$ (methane, 2900 – 3200 cm$^{-1}$), $C_2H_6$ (ethane, around 3000 cm$^{-1}$), and $C_2H_2$ (acetylene, around 3300 cm$^{-1}$) in 1 atmosphere of nitrogen ($N_2$) with a bandwidth of 650 cm$^{-1}$ and 110,000 comb modes. Both transmittance and phase response are shown along with the Hitran 2008 [36] model (red line) . Over the expanded view, the transmission noise is 0.002 and the phase noise is 2 mrad.

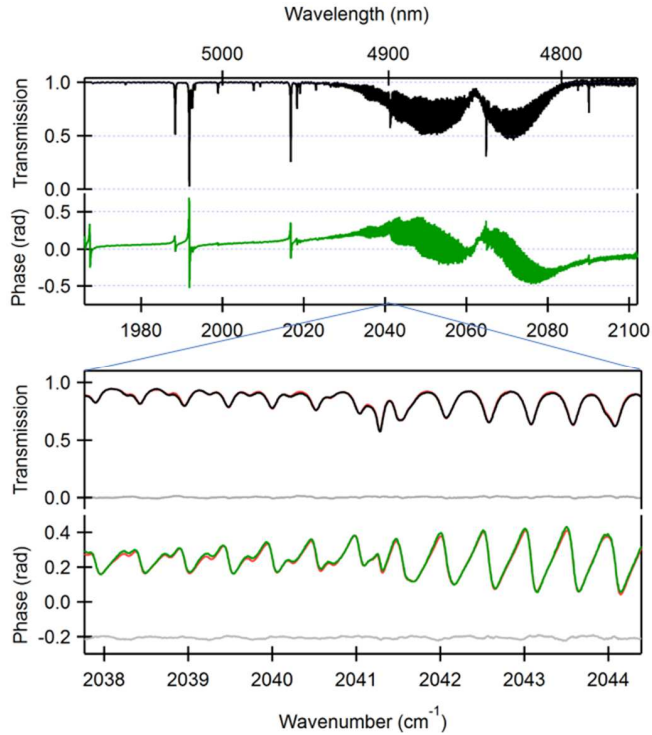

Figure 5. **Measured spectrum of carbonyl sulfide** (COS) in 1 atmosphere of nitrogen ($N_2$) acquired over 80 minutes. Both transmittance and phase response are shown along with the Hitran 2008 [36] model (red line) and residuals (gray line, phase residuals offset by -0.2 rad). The spikes are $H_2O$ lines. At 2000 $cm^{-1}$, the phase noise is 200 µrad; over the shown spectral range it averages to 4.5 mrad.

Spectra were coherently averaged for 32 minutes, achieving a peak spectral SNR – defined as the ratio of the signal to the random variations of the baseline -- of 660 or 15 $/\sqrt{s}$. The overall comb intensity structure was removed using standard baseline fitting techniques to find the transmission and phase spectra shown, which agree well with a model based on Hitran 2008 line parameters and Voigt lineshapes [36]. Given the excellent agreement of the mid-infrared DCS with the narrow, well calibrated Hydrogen cyanide (HCN) lines shown in Fig. 2c, we attribute any differences to the Hitran 2008 database, illustrating the potential of mid-infrared DCS to contribute to improved spectral databases.

Figure 5 shows the measured spectrum at longer wavelengths from 5.08 µm to 4.75 µm, corresponding to a bandwidth of 136 $cm^{-1}$ (4.1 THz) and 20,000 resolved comb modes. Spectra were coherently averaged for 80 minutes, achieving a peak spectral SNR of 6500, or 94 $/\sqrt{s}$. As before, the phase and transmission spectra were extracted after standard baseline correction for the comb structure and agree very well with the Hitran 2008 based model.

We can compare different spectra and dual-comb systems through a DCS figure-of-merit [37], defined as the average SNR (not peak) times the number of comb modes, which is then identical to simply summing the SNR across the comb modes (see Methods). The SNR refers to the uncorrelated point-to-point spectral fluctuations and ignores slower variations from, for example, any changing etalons in the beam paths. We find a figure-of-merit of $1\times10^6/\sqrt{s}$ for the propane spectrum of Fig. 3 and $6\times10^5/\sqrt{s}$ for both the 3-µm Figure 4 (b) and the 5-µm spectrum of Figure 5 (a). The figures-of-merit are limited by the relative intensity noise (RIN) on the mid-infrared combs of ~ -127 dBc/Hz. In the future, the use of a balanced HgCdTe detector could suppress this noise substantially as in the near-infrared [38]. While these figures-of-merit do not reach the shot-noise limit, they are comparable to reported results for narrow-band DFG DCS, which range from $1.0 - 2.3\times10^6/\sqrt{s}$ [13,20].

### Broad instantaneous bandwidth through waveguide PPLN

Even broader instantaneous bandwidths are interesting for many applications, for example for covering the full 3—4 µm and 4.5—5 µm atmospheric windows, but are challenging to achieve in bulk PPLN crystals. To extend our interaction length while maintaining high intensities, we designed an aperiodically-poled lithium niobate ridge waveguide. As shown in Fig. 6a, the waveguide is 15 µm×15 µm in cross section and 25 mm in length on a lithium tantalate substrate. The poling along the 25 mm-long waveguide is varied from periods of 22 to 30 µm to achieve phase matching for DFG from 2.6 to 5.2 µm. To mitigate the tendency of optical parametric amplification to effectively reduce the difference-frequency bandwidth, the poling period is chirped at a rate proportional to the phase matching bandwidth, as shown in Fig. 6a. Qualitatively, less waveguide length is devoted to poling periods with broad phase-matching bandwidth and *vice versa*. To generate a broad-bandwidth mid infrared spectrum, we stretch our pump pulse to ~1 ps before coupling into the waveguide, with the final mid-IR spectrum dependent on the relative delay between the input pump and signal light. In addition to the broad bandwidth, the waveguide PPLN outputs a mid-IR beam of higher spatial quality than the bulk crystal, which facilitates coupling into single-mode fiber. Supplemental Figure 2 shows one example of a smooth spectrum covering 3.2-3.7 μm at 90 mW average power, which is coupled to single-mode ZrF$_4$ fiber with 30 % efficiency.

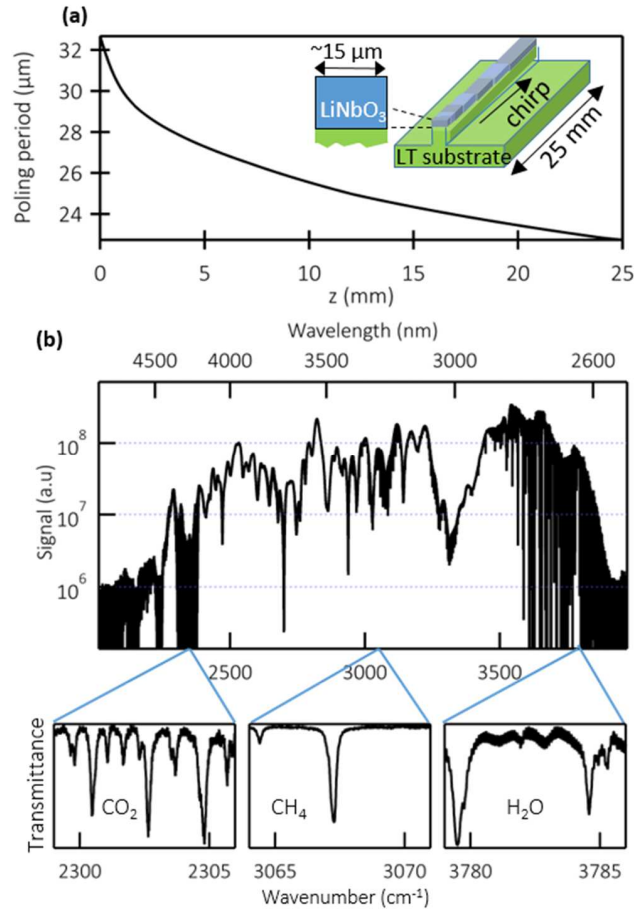

Figure 6 **Broad-band dual-comb spectroscopy with light generated in PPLN waveguides. (a)** Aperiodic poling design of lithium niobate waveguides used, showing variation of poling periods from waveguide entrance (z=0) to exit (z=25 mm). Inset: waveguide dimensions. **(b)** Broadband dual-comb spectrum obtained with two chirped waveguide PPLNs. The covered span is 1640 cm$^{-1}$ (2250 – 3890 cm$^{-1}$), corresponding to 245,000 teeth at 200 MHz resolution. The bottom panels show the molecular spectra after performing a baseline correction; the broad spectral span covered within one single measurement allows to simultaneously detect multiple species, as shown for $CO_2$, $CH_4$, and $H_2O$.

Figure 6b shows a DCS spectrum acquired using mid-infrared combs, each generated from chirped waveguide PPLN, for a gas cell containing $CO_2$, $CH_4$ and $H_2O$. While the spectra produced by the waveguide PPLN are more structured than those produced by the bulk PPLN, they are broader. The DCS spectrum covers 2250 – 3890 cm$^{-1}$, corresponding to a bandwidth of 1640 cm$^{-1}$ (1874 nm and 49 THz) and 245,000 comb modes. Equally important, there is no evidence of higher intensity noise; averaging for 83 minutes, the spectrum achieves a peak spectral SNR of 2000 (28 $/\sqrt{s}$) and a figure of merit equal to $9\times10^5 / \sqrt{s}$. With further design, the chirped waveguide PPLNs should be able to provide high-power, single spatial mode mid-infrared light with spectral coverage tailored to different broadband applications.

In conclusion, the system shown here is capable of comb-tooth resolved, broadband spectroscopy in the functional group index region from 3—5 microns with high signal-to-noise ratio and with frequency resolution and accuracy equal to or exceeding high performance mid-infrared FTS but at 75x faster acquisition time. In contrast to FTS, this performance is reached with a high power, bright laser source (attenuated here to avoid detector saturation) thereby enabling detection over long, lossy paths. It is also realized in a relatively compact optical setup without a long mechanical delay line, enabling future portable applications. With these combined attributes, mid-infrared DCS has the potential for a much broader set of non-laboratory applications than conventional high-resolution FTS. We envision applications ranging from industrial process monitoring to outdoor atmospheric measurements, where the broadband mid-infrared coverage could enable sensitive, rapid detection of large molecules even in the presence of strongly absorbing background gases.

We acknowledge funding from the Defense Advanced Research Projects Agency Defense Sciences Office SCOUT program, discussions with Flavio Cruz regarding the design of the lithium niobate waveguides, and helpful comments from Adam Fleisher and Eleanor Waxman.

Contributions

The experiments were conceived of by N.R.N., I.C., G.Y. and S.A.D. The mid-infrared systems were built by G.Y., E.B., and D.H. The digital signal processing was implemented by G.Y. and F.R.G. Data analysis was performed by F.R.G. and G.Y. The manuscript was written by G.Y., F.R.G., I.C., and N.R.N.

**REFERENCES**

1. I. Coddington, N. Newbury, and W. Swann, "Dual-comb spectroscopy," Optica **3**, 414 (2016).
2. T. Ideguchi, "Dual-Comb Spectroscopy," Opt. Photonics News **28**, 32 (2017).
3. S. Boudreau, S. Levasseur, C. Perilla, S. Roy, and J. Genest, "Chemical detection with hyperspectral lidar using dual frequency combs," Opt. Express **21**, 7411–7418 (2013).
4. P. J. Schroeder, R. J. Wright, S. Coburn, B. Sodergren, K. C. Cossel, S. Droste, G. W. Truong, E. Baumann, F. R. Giorgetta, I. Coddington, N. R. Newbury, and G. B. Rieker, "Dual frequency comb laser absorption spectroscopy in a 16 MW gas turbine exhaust," Proc. Combust. Inst. 10.1016/j.proci.2016.06.032 (2016).
5. E. M. Waxman, K. C. Cossel, G.-W. Truong, F. R. Giorgetta, W. C. Swann, S. C. Coburn, R. J. Wright, G. B. Rieker, I. Coddington, and N. R. Newbury, "Comparison of Open-Path Dual Frequency Comb Spectroscopy for High-Precision Atmospheric Gas Measurements," Atmos Meas Tech Discuss **in review**, doi: 10.5194/amt-2017-62 (2017).
6. K. C. Cossel, E. M. Waxman, F. R. Giorgetta, M. Cermak, I. R. Coddington, D. Hesselius, S. Ruben, W. C. Swann, G.-W. Truong, G. B. Rieker, and N. R. Newbury, "Open-path dual-comb spectroscopy to an airborne retroreflector," Optica **4**, 724–728 (2017).
7. J. Roy, J.-D. Deschênes, S. Potvin, and J. Genest, "Continuous real-time correction and averaging for frequency comb interferometry," Opt. Express **20**, 21932–21939 (2012).
8. A. M. Zolot, F. R. Giorgetta, E. Baumann, J. W. Nicholson, W. C. Swann, I. Coddington, and N. R. Newbury, "Direct-comb molecular spectroscopy with accurate, resolved comb teeth over 43 THz," Opt. Lett. **37**, 638–640 (2012).
9. T. Ideguchi, A. Poisson, G. Guelachvili, N. Picqué, and T. W. Hänsch, "Adaptive real-time dual-comb spectroscopy," Nat. Commun. **5**, 3375 (2014).
10. S. Okubo, K. Iwakuni, H. Inaba, K. Hosaka, A. Onae, H. Sasada, and F.-L. Hong, "Ultra-broadband dual-comb spectroscopy across 1.0–1.9 μm," Appl. Phys. Express **8**, 082402 (2015).
11. A. Schliesser, M. Brehm, F. Keilmann, and D. van der Weide, "Frequency-comb infrared spectrometer for rapid, remote chemical sensing," Opt. Express **13**, 9029–9038 (2005).
12. B. Bernhardt, E. Sorokin, P. Jacquet, R. Thon, T. Becker, I. T. Sorokina, N. Picqué, and T. W. Hänsch, "Mid-infrared dual-comb spectroscopy with 2.4 μm Cr2+:ZnSe femtosecond lasers," Appl. Phys. B **100**, 3–8 (2010).
13. E. Baumann, F. R. Giorgetta, W. C. Swann, A. M. Zolot, I. Coddington, and N. R. Newbury, "Spectroscopy of the methane $v_3$ band with an accurate midinfrared coherent dual-comb spectrometer," Phys. Rev. A **84**, 062513 (2011).
14. A. Schliesser, N. Picqué, and T. W. Hänsch, "Mid-infrared frequency combs," Nat. Photonics **6**, 440–449 (2012).
15. Z. Zhang, T. Gardiner, and D. T. Reid, "Mid-infrared dual-comb spectroscopy with an optical parametric oscillator," Opt. Lett. **38**, 3148–3150 (2013).
16. G. Villares, A. Hugi, S. Blaser, and J. Faist, "Dual-comb spectroscopy based on quantum-cascade-laser frequency combs," Nat. Commun. **5**, 5192 (2014).
17. F. Zhu, A. Bicer, R. Askar, J. Bounds, A. A. Kolomenskii, V. Kelessides, M. Amani, and H. A. Schuessler, "Mid-infrared dual frequency comb spectroscopy based on fiber lasers for the detection of methane in ambient air," Laser Phys. Lett. **12**, 095701 (2015).

18. Y. Jin, S. M. Cristescu, F. J. M. Harren, and J. Mandon, "Femtosecond optical parametric oscillators toward real-time dual-comb spectroscopy," Appl. Phys. B **119**, 1–10 (2015).
19. F. C. Cruz, D. L. Maser, T. Johnson, G. Ycas, A. Klose, F. R. Giorgetta, I. Coddington, and S. A. Diddams, "Mid-infrared optical frequency combs based on difference frequency generation for molecular spectroscopy," Opt. Express **23**, 26814 (2015).
20. M. Yan, P.-L. Luo, K. Iwakuni, G. Millot, T. W. Hänsch, and N. Picqué, "Mid-infrared dual-comb spectroscopy with electro-optic modulators," Light Sci. Appl. Accept. Artic. Preview 3 (2017).
21. M. Yu, Y. Okawachi, A. G. Griffith, N. Picqué, M. Lipson, and A. L. Gaeta, "Silicon-chip-based mid-infrared dual-comb spectroscopy," ArXiv161001121 Phys. (2016).
22. V. O. Smolski, H. Yang, J. Xu, and K. L. Vodopyanov, "Massively parallel dual-comb molecular detection with subharmonic optical parametric oscillators," ArXiv160807318 Phys. (2016).
23. J. Westberg, L. A. Sterczewski, and G. Wysocki, "Mid-infrared multiheterodyne spectroscopy with phase-locked quantum cascade lasers," Appl. Phys. Lett. **110**, 141108 (2017).
24. O. Kara, Z. Zhang, T. Gardiner, and D. T. Reid, "Dual-comb mid-infrared spectroscopy with free-running oscillators and absolute optical calibration from a radio-frequency reference," Opt. Express **25**, 16072 (2017).
25. L. C. Sinclair, J.-D. Deschênes, L. Sonderhouse, W. C. Swann, I. H. Khader, E. Baumann, N. R. Newbury, and I. Coddington, "Invited Article: A compact optically coherent fiber frequency comb," Rev. Sci. Instrum. **86**, 081301 (2015).
26. C. Erny, K. Moutzouris, J. Biegert, D. Kühlke, F. Adler, A. Leitenstorfer, and U. Keller, "Mid-infrared difference-frequency generation of ultrashort pulses tunable between 3.2 and 4.8 µm from a compact fiber source," Opt. Lett. **32**, 1138–1140 (2007).
27. D. L. Maser, G. Ycas, W. I. Depetri, F. C. Cruz, and S. A. Diddams, "Coherent frequency combs for spectroscopy across the 3?5 ?m region," Appl. Phys. B **123**, (2017).
28. G. Villares, J. Wolf, D. Kazakov, M. J. Süess, A. Hugi, M. Beck, and J. Faist, "On-chip dual-comb based on quantum cascade laser frequency combs," Appl. Phys. Lett. **107**, 251104 (2015).
29. J.-D. Deschênes, P. Giaccarri, and J. Genest, "Optical referencing technique with CW lasers as intermediate oscillators for continuous full delay range frequency comb interferometry," Opt. Express **18**, 23358–23370 (2010).
30. I. Coddington, W. C. Swann, and N. R. Newbury, "Coherent dual-comb spectroscopy at high signal-to-noise ratio," Phys Rev A **82**, 043817 (2010).
31. M. Jacobsen, D. Richmond, M. Hogains, and R. Kastner, "RIFFA 2.1: A Reusable Integration Framework for FPGA Accelerators," ACM Trans. Reconfigurable Technol. Syst. **8**, 1–23 (2015).
32. V. Malathy Devi, D. C. Benner, M. A. H. Smith, C. P. Rinsland, S. W. Sharpe, and R. L. Sams, "A multispectrum analysis of the v1 band of H12C14N: Part I. Intensities, self-broadening and self-shift coefficients," J. Quant. Spectrosc. Radiat. Transf. **82**, 319–341 (2003).
33. C. P. Rinsland, V. Malathy Devi, M. A. H. Smith, D. Chris Benner, S. W. Sharpe, and R. L. Sams, "A multispectrum analysis of the v1 band of H12C14N: Part II. Air- and N2-broadening, shifts and their temperature dependences," J. Quant. Spectrosc. Radiat. Transf. **82**, 343–362 (2003).
34. C. W. Moore, B. Zielinska, G. Pétron, and R. B. Jackson, "Air Impacts of Increased Natural Gas Acquisition, Processing, and Use: A Critical Review," Environ. Sci. Technol. **48**, 8349–8359 (2014).
35. C. A. Beale, R. J. Hargreaves, and P. F. Bernath, "Temperature-dependent high resolution absorption cross sections of propane," J. Quant. Spectrosc. Radiat. Transf. **182**, 219–224 (2016).
36. L. S. Rothman, I. E. Gordon, A. Barbe, D. C. Benner, P. E. Bernath, M. Birk, V. Boudon, L. R. Brown, A. Campargue, J. P. Champion, K. Chance, L. H. Coudert, V. Dana, V. M. Devi, S. Fally, J. M. Flaud, R. R. Gamache, A. Goldman, D. Jacquemart, I. Kleiner, N. Lacome, W. J. Lafferty, J. Y. Mandin, S. T. Massie, S. N. Mikhailenko, C. E. Miller, N. Moazzen-Ahmadi, O. V. Naumenko, A. V. Nikitin, J. Orphal, V. I. Perevalov, A. Perrin, A. Predoi-Cross, C. P. Rinsland, M. Rotger, M. Simeckova, M. A. H. Smith, K. Sung, S. A. Tashkun, J. Tennyson, R. A. Toth, A. C. Vandaele, and J. Vander Auwera, "The HITRAN 2008 molecular spectroscopic database," J. Quant. Spectrosc. Radiat. Transf. **110**, 533–572 (2009).
37. N. R. Newbury, I. Coddington, and W. C. Swann, "Sensitivity of coherent dual-comb spectroscopy," Opt. Express **18**, 7929–7945 (2010).
38. G.-W. Truong, E. M. Waxman, K. C. Cossel, E. Baumann, A. Klose, F. R. Giorgetta, W. C. Swann, N. R. Newbury, and I. Coddington, "Accurate frequency referencing for fieldable dual-comb spectroscopy," Opt. Express **24**, 30495–30504 (2016).

# Octave-spanning Mid-Infrared Dual Comb Spectroscopy with Comb-Tooth Resolution and High Signal-To-Noise Ratio: Methods

GABRIEL YCAS[1*], FABRIZIO R. GIORGETTA[1], ESTHER BAUMANN[1], IAN CODDINGTON[1], DANIEL HERMAN[1], SCOTT A. DIDDAMS[1], NATHAN R. NEWBURY[1]

[1]*NIST 325 Broadway, Boulder, CO 80305*
*Corresponding author: ycasg@colorado.edu*

**Coherent Averaging**

A dual-comb spectrometer using coherent averaging should exhibit a signal-to-noise ratio that increases with the square-root of the number of averages. We define the noise as the "spectral noise", or the standard deviation of the dual-comb spectrum after broad features are removed by high-pass filtering the Fourier-transformed interferogram.

**Figure-of-Merit**

We calculate the figure-of-merit [1] by extending the expression

(S6) $$FOM = m \times SNR / \sqrt{t}$$

so that we can account for the variation in signal-to-noise across the spectrum. Our figure-of-merit averages the SNR across the modes and is defined as

(S7) $$FOM = \sum_{n=n_{low}}^{n_{high}} \frac{I_n}{\sqrt{t}\sigma_n},$$

where $n_{low}$ and $n_{high}$ are the indices of the lowest and highest frequency spectral elements in the dual-comb spectrum, $I_n$ is the signal strength of each element, and $\sigma_n$ is the one-sigma deviation at mode $n$.

In one of our measurements we recorded both 600 ms of continuous data as well as averaged data of up to 100,000 interferograms, recorded as ten consecutive sets of 10,000 averages each. From this data, we calculate the spectral SNR as a function of averaging time over 5 orders of magnitude. As shown in Fig. S1, the peak SNR increases as a function of $\sqrt{N}$ as expected, in this case increasing by 8.1 $/\sqrt{N}$.

**Alternate configuration of Waveguides**

While producing broad-band optical spectra from the aperiodically poled lithium niobate waveguides it became apparent that our 25 mm-long waveguides provided an interaction length that was too long for the generation of smooth optical spectra. We therefore cut one of our waveguides, keeping just the first 10 mm with poling period ranging from 32-26 microns. Using this short waveguide, a DFG spectrum that is smooth and with 90 mW of average power could be generated, shown in Fig. S2. Using a 15 mm focal length off-axis paraboloid (OAP) mirror to collimate the light leaving the waveguide and a 15 mm focal length OAP mirror to focus light into a single-mode ZrF4 fiber, a maximum of 30 mW were coupled into the fiber.

## References

1.    N. R. Newbury, I. Coddington, and W. C. Swann, "Sensitivity of coherent dual-comb spectroscopy," Opt. Express **18**, 7929–7945 (2010).

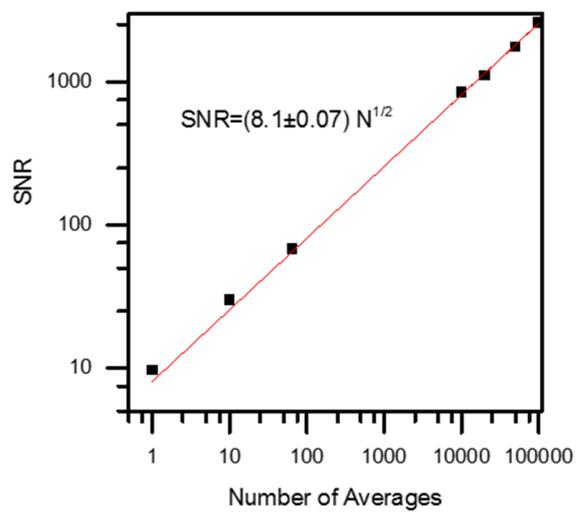

Figure S1 Peak spectral signal-to-noise of dual-comb spectra recorded around 4.8 microns as function of number of averages.

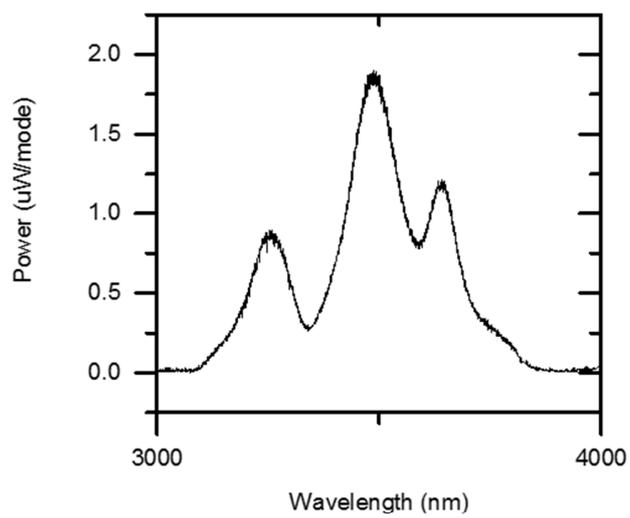

Figure S2 Optical spectrum from 10 mm-long waveguide segment measured with a Fourier transform spectrometer. It is optimized for producing light between 3.2-3.7 um. Of the 90 mW produced by the waveguide, 30 mW were coupled into single-mode ZrF4 fiber.